\documentclass[10pt]{article}
\usepackage[utf8]{inputenc}
\usepackage[square,numbers,sort&compress]{natbib}
\usepackage{authblk}
\usepackage{amsmath,amssymb,physics,bm}
\usepackage{graphicx}
\usepackage[top=1 in,bottom=1in, left=1 in, right=1 in]{geometry}
\usepackage{setspace}
\usepackage{xcolor}

\title{\Large{\bf{Incident-polarization-independent spin Hall effect of light reaching a half beam waist}}}
\author[1]{Minkyung Kim}
\author[2]{Dasol Lee}
\author[1,3,4,*]{Junsuk Rho}
\affil[1]{\small{Department of Mechanical Engineering, Pohang University of Science and Technology (POSTECH), Pohang 37673, Republic of Korea}}
\affil[2]{\small{Department of Biomedical Engineering, Yonsei University, Wonju 26493, Republic of Korea}}
\affil[3]{\small{Department of Chemical Engineering, Pohang University of Science and Technology (POSTECH), Pohang 37673, Republic of Korea}}
\affil[4]{\small{POSCO-POSTECH-RIST Convergence Research Center for Flat Optics and Metaphotonics, Pohang 37673, Republic of Korea}}
\affil[*]{\small{jsrho@postech.ac.kr}}
\date{}

\renewcommand{\vec}[1]{\mathbf{#1}}

\begin{document}

\maketitle

\begin{abstract}
The spin Hall effect of light, a spin-dependent transverse splitting of light at an optical interface, is intrinsically an incident-polarization-sensitive phenomenon. Recently, an approach to eliminate the polarization dependence by equalizing the reflection coefficients of two linear polarizations has been proposed, but is only valid when the beam waist is sufficiently larger than the wavelength. Here, we demonstrate that an interface, at which the reflection coefficients of the two linear polarizations are the same and so are their derivatives with respect to the incident angle, supports the polarization-independent spin Hall shift, even when the beam waist is comparable to the wavelength. In addition, an isotropic-anisotropic interface that exhibits the polarization-independent spin Hall shift over the entire range of incident angles is presented. Monte-Carlo simulations prove that spin Hall shifts are degenerate under any polarization and reaches a half of beam waist under unpolarized incidence. We suggest an application of the beam-waist-scale spin Hall effect of light as a tunable beam-splitting device that is responsive to the incident polarization. The spin Hall shift that is independent of the incident polarization at any incident angle will facilitate a wide range of applications including practical spin-dependent devices and active beam splitters.
\end{abstract}

\section{Introduction}
The spin Hall effect of light (SHEL) \cite{onoda2004hall, hosten2008observation, PhysRevD.5.787, fedorov2013theory, Ling_2017}, a spin-dependent microscopic splitting of refracted/reflected light along the transverse direction, is an optical analog of the transverse spin accumulation of electric current, called the spin Hall effect \cite{dyakonov1971current, d1971possibility}. A linearly polarized incidence is split into left circularly polarized light (LCP) and right circularly polarized light (RCP) that are displaced in opposite directions. The origin of the SHEL lies on a finite beam waist of the incidence that contains wave vectors with slightly different orientations and on the transversality of light \cite{hosten2008observation, Bliokh_2013, bliokh2015spin}, which enforces orthogonal locking between the polarization state and momentum. Wave vectors that have nonzero components along the transverse direction yield local rotations of the polarization basis, resulting in a spin-dependent position correction term in the refracted/reflected beam, perpendicular to the incident plane. Therefore, the SHEL is inherently dependent on incident polarization.

\begin{figure}[h!] \centering
	\includegraphics[width = 0.5\textwidth]{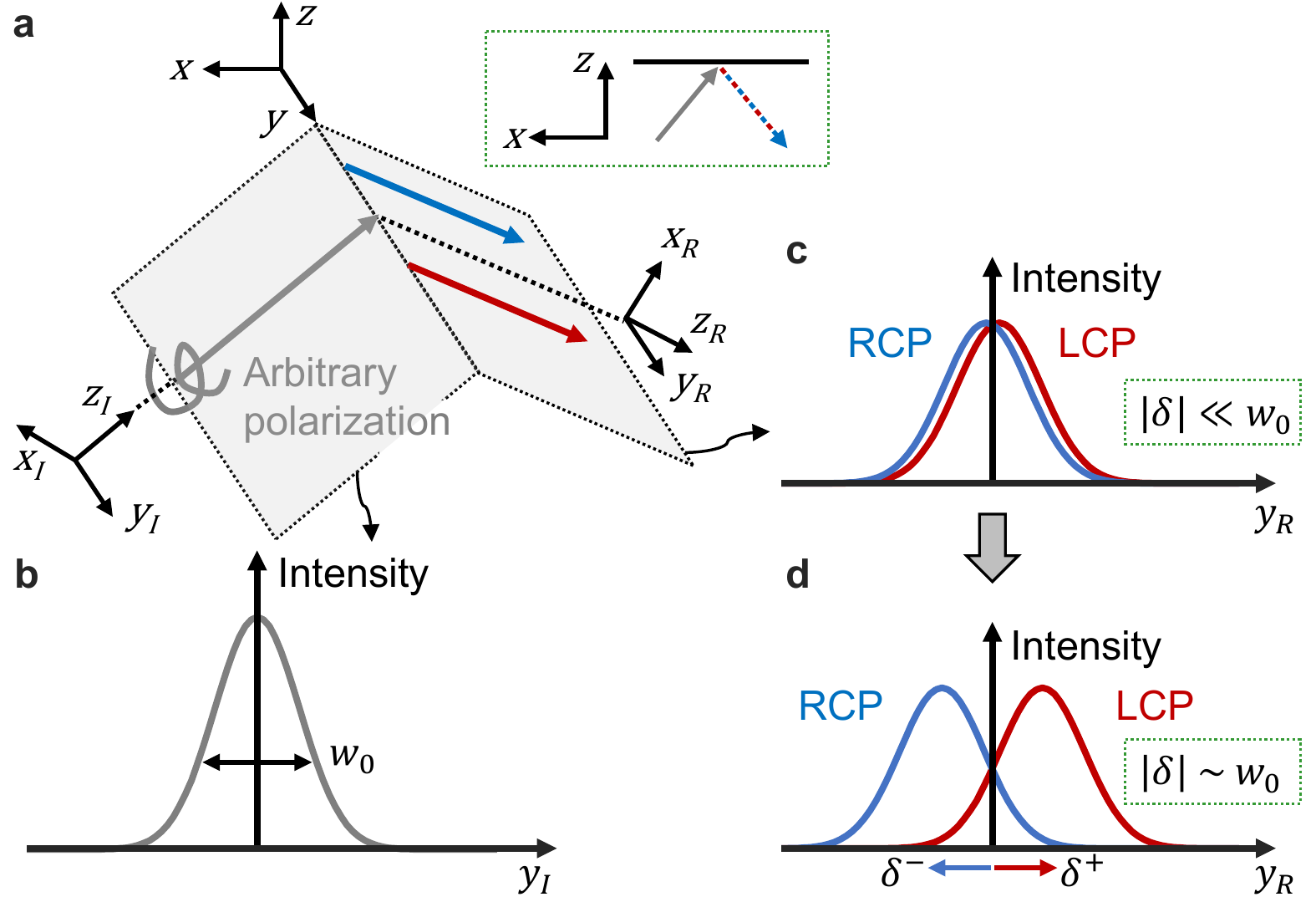}
	\caption{Schematic of the polarization-independent SHEL. (a) The SHEL of a reflected beam at an interface ($z = 0$). Inset shows a two-dimensional view in the incident plane. (b) Intensity profile of the incidence along the transverse direction. (c, d) Intensity profiles of the circularly polarized components of the reflected beam along the transverse direction when the spin Hall shift ($\delta$) is (c) much smaller than or (d) comparable to the beam waist.}
	\label{schematic}
\end{figure}

It has been reported recently that an interface that has the same reflection coefficients, i.e., $r_s = r_p$, where subscripts $s$ and $p$ denote incident polarization states, supports the incident-polarization-independent SHEL for the reflected beam, or equivalently transmission coefficients $t_s = -t_p$ for the refracted beam \cite{https://doi.org/10.1002/lpor.202100138}. The spin Hall shifts at such an interface are degenerate under incidences that have any polarization state (Fig. \ref{schematic}a, b). One assumption that underlies the theoretical demonstration is that the beam waist $w_0$ is much larger than wavelength $\lambda$, satisfying
\begin{equation}
	w_0^2 \gg \Big(\frac{\lambda \cot\theta_i}{2\pi} \Big)^2,
	\label{large_beam}
\end{equation}	
where $\theta_i$ is the incident angle of the beam. We refer to this condition as a large beam waist condition hereafter for convenience. While this condition is valid in most instances except for tightly-confined or near-normal incidence, the spin Hall shift under the condition is essentially much smaller than $w_0$ (Fig. \ref{schematic}c) and breaking of the condition provides a route to increase the spin Hall shift substantially even up to $w_0/2$. Given that tremendous efforts have been devoted during the past decade to increasing the spin Hall shift from the deep-subwavelength scale to several wavelengths and even beyond \cite{PhysRevA.84.043806, yin2013photonic, Zhu:15, 7394110, 7970110, Jiang_2018, Jiang_2018_BIC, Takayama:18, kim2019observation, PhysRevLett.124.053902, doi:10.1063/5.0009616, https://doi.org/10.1002/lpor.202000393, yang2021dynamic, doi:10.1021/acsphotonics.1c00727}, an approach to achieve polarization-independent SHEL that is valid even when $w_0^2$ is comparable to or less than $(\lambda \cot\theta_i/2\pi)^2$ is in high demand (Fig. \ref{schematic}d).

In this article, we demonstrate a gigantic SHEL that reaches $w_0/2$ and is insensitive to the incident polarization at an interface between isotropic and anisotropic media. The dependence of the spin Hall shift on incident polarization is removed even for light that does not satisfy the large beam waist condition (Eq. \ref{large_beam}) if the Fresnel coefficients of two linear polarizations as well as their angular derivatives are equal to each other. Starting from the theoretical description to prove the polarization independence under such conditions, we demonstrate both analytically and numerically that such conditions are satisfied in the entire range of $0 < \theta_i < \pi$ if anisotropic and isotropic media adjacent to each other have permittivities that follow a certain relation. Furthermore, Monte-Carlo simulations are conducted to confirm the degenerate SHEL under arbitrarily polarized incidence. The spatial distributions of the reflected beam clearly manifest a distinguishable splitting that reaches $w_0/2$ under unpolarized incidence. This gigantic SHEL that is insensitive to the incident polarization at any $\theta_i$ opens a path towards realization of spin-sensitive and compact devices that are robust against change of incident polarization and dynamic beam splitters that operate in response to the incident polarization.

\section{Conditions for the polarization-independent SHEL}
This section revisits the derivation of the reflected beam profile and the spin Hall shift using a wave packet model and deduces the required conditions for polarization-independent SHEL that do not require any assumption or condition (Eq. \ref{large_beam}). In a macroscopic view, the refraction and reflection at a planar interface are described by Snell's law and Fresnel equations. However, those equations assume a single wave vector for each incident, refracted, and reflected beam and thus are not correct microscopically if the beams have a finite beam waist. More specifically, an incidence not only has the central wave vector parallel to the $k_{Iz}$-axis, but also contains noncentral wave vectors that have nonzero $k_{Ix}$ and $k_{Iy}$ components (Fig. \ref{schematic}a). Thus, two correction terms should be added, one for $k_{Ix}$ and the other for $k_{Iy}$. First, to consider nonzero $k_{Iy}$, the Jones matrix at the interface has off-diagonal terms \cite{PhysRevA.84.043806}
\begin{equation}
	\begin{pmatrix} \psi^H_R \\ \psi^V_R \end{pmatrix} = \begin{pmatrix} r_p & \frac{k_y}{k_0} (r_p + r_s) \cot{\theta_i} \\ -\frac{k_y}{k_0} (r_p + r_s) \cot{\theta_i} & r_s \end{pmatrix} \begin{pmatrix} \psi^H_I \\ \psi^V_I \end{pmatrix},
	\label{jones}
\end{equation}
where superscripts $H$ and $V$ denote horizontal and vertical polarizations respectively, subscripts $I$ and $R$ correspond to the incident and reflected beam respectively, $k_0$ is the incident wave vector, $k_x \equiv k_{Ix} = -k_{Rx}$ and $k_y \equiv k_{Iy}  = k_{Ry}$ are $x$- and $y$-components of the wave vector, and the incident and reflected beams have a Gaussian beam shape as
\begin{equation}
	\psi_{I, R} = \begin{pmatrix} \psi_{I,R}^H \\ \psi_{I,R}^V \end{pmatrix} \frac{w_0}{\sqrt{2\pi}} \exp(-\frac{w_0 (k_x^2 + k_y^2)}{4}).
\end{equation}
Then, the nonzero $k_{Ix}$ is considered by applying a Taylor series expansion to $r_s$ and $r_p$ in Eq. \ref{jones} as
\begin{equation}
	r_{s,p} = r_{s,p}(k_{Ix} = 0) + \frac{\partial r_{s,p}}{\partial k_{Ix}} \bigg\rvert_{k_{Ix} = 0} k_{Ix}.
	\label{taylor}
\end{equation}
Combining Eqs. \ref{jones} to \ref{taylor} and applying basis transformation from linear to circular and an inverse Fourier transform provide the spatial field profiles of circularly polarized components of the reflected beam as linear equations $\tilde{\psi}^{\pm}_R = \tilde{\psi}^\pm_{R,H} \psi^{H}_I \pm i \tilde{\psi}^\pm_{R,V} \psi^{V}_I$ where
\begin{align}
	\tilde{\psi}^\pm_{R,H} =& \frac{1}{\sqrt{2\pi} w_0} \frac{z_0}{z_0 + i z_R} \exp(-\frac{k_0}{2}  \frac{x_R^2+y_R^2}{z_0 + i z_R}) \notag \\
	&\times \Big[ r_p  - i \frac{x_R}{z_0 + i z_R} \dot{r_p} \mp \frac{y_R \cot{\theta_i}}{z_0 + i z_R} (r_p + r_s) \mp i \frac{x_R y_R \cot{\theta_i}}{(z_0 + i z_R)^2} (\dot{r_p} + \dot{r_s}) \Big] \exp(ik_r z_R), \notag \\
	\tilde{\psi}^\pm_{R,V} =& \frac{1}{\sqrt{2\pi} w_0} \frac{z_0}{z_0 + i z_R} \exp(-\frac{k_0}{2}  \frac{x_R^2+y_R^2}{z_0 + i z_R}) \notag \\
	&\times \Big[ r_s  - i \frac{x_R}{z_0 + i z_R} \dot{r_s} \mp \frac{y_R \cot{\theta_i}}{z_0 + i z_R} (r_p + r_s) \mp i \frac{x_R y_R \cot{\theta_i}}{(z_0 + i z_R)^2} (\dot{r_p} + \dot{r_s}) \Big] \exp(ik_r z_R),
	\label{fieldHV}
\end{align}
where the second subscript of $\tilde{\psi}$ indicates incident polarization, the dot notation represents the derivative with respect to $\theta_i$ ($\dot{r}_{s,p} = \partial r_{s,p} /\partial \theta_i$), and $z_0 = k_0 w_0^2/2$ is the Rayleigh length. Taking the $y$ position average 
\begin{equation}
	\delta^{\pm} = \frac{\expval{y_R}{\tilde{\psi}_R^{\pm}}}{\bra{\tilde{\psi}_R^{\pm}}\ket{\tilde{\psi}_R^{\pm}}}
	\label{yaverage}
\end{equation}
of the reflected beam provides the exact formula of the spin Hall shift \cite{miao2021limitations}:
\begin{align}
	\delta^\pm_H = \mp \frac{\cot{\theta_i}}{k_0} \frac{\lvert r_p \rvert ^2 + \text{Re}(r_p r_s^*)}{\lvert r_p \rvert ^2 + (\frac{\cot{\theta_i}}{k_0 w_0})^2 \lvert r_p + r_s \rvert^2 + \frac{1}{k_0^2 w_0^2}\dot{r}_p \dot{r}_p^*}, \notag \\
	\delta^\pm_V = \mp \frac{\cot{\theta_i}}{k_0} \frac{\lvert r_s \rvert ^2 + \text{Re}(r_p^* r_s)}{\lvert r_s \rvert ^2 + (\frac{\cot{\theta_i}}{k_0 w_0})^2 \lvert r_p + r_s \rvert^2 + \frac{1}{k_0^2 w_0^2}\dot{r}_s \dot{r}_s^*},
	\label{shift_exact}
\end{align}
where $*$ corresponds to the complex conjugate. Eq. \ref{shift_exact} can be further simplified to
\begin{align}
	\delta_H^\pm &= \mp \frac{\cot{\theta_i}}{k_0} \text{Re} (1 + \frac{r_s}{r_p}), \notag \\
	\delta_V^\pm &= \mp \frac{\cot{\theta_i}}{k_0} \text{Re} (1 + \frac{r_p}{r_s}),
	\label{shift}
\end{align}
under the large beam waist condition (Eq. \ref{large_beam}).

A close examination of Eq. \ref{fieldHV} shows that the two field profiles appear symmetrical ($r_s \rightarrow r_p$, $r_p \rightarrow r_s$, $\dot{r}_s \rightarrow \dot{r}_p$, and $\dot{r}_p \rightarrow \dot{r}_s$). Therefore, if $r_s = r_p$ and $\dot{r}_s = \dot{r}_p$, then $\tilde{\psi}^\pm_{R,H} = \tilde{\psi}^\pm_{R,V} \equiv \tilde{\psi}^\pm_{R0}$ and thus the components of the incident Jones vector $\psi_I^H$ and $\psi_I^V$ act only as constant coefficients of the field profile as $\tilde{\psi}^{\pm}_R = (\psi^{H}_I \pm i \psi^{V}_I) \tilde{\psi}^\pm_{R0}$. Then substitution of this equation to Eq. \ref{yaverage} provides $\delta^{\pm} = \expval{y_R}{\tilde{\psi}_{R0}^{\pm}}/\bra{\tilde{\psi}_{R0}^{\pm}}\ket{\tilde{\psi}_{R0}^{\pm}}$, which contains neither $\psi_I^H$ nor $\psi_I^V$. Therefore, the spin Hall shift is independent of the polarization state of the incidence. Similarly, the polarization-independent SHEL can be also demonstrated for the transmitted beam if $t_s = -t_p$ and $\dot{t}_s = -\dot{t}_p$.

Importantly, although the previous condition of $r_s = r_p$ is obtained under the assumption that the wave vector deflections are much smaller than the wave number ($k_{x,y} \ll k_0$) and is valid under the large beam waist condition (Eq. \ref{large_beam}) \cite{https://doi.org/10.1002/lpor.202100138}, the theoretical proof here does not rely on any assumption. Thus, the SHEL at an interface that supports $r_s = r_p$ and $\dot{r}_s = \dot{r}_p$ appears independent of the incident polarization, under any condition, even when $w_0$ is comparable to $\lambda$ or $\theta_i$ is so small that Eq. \ref{large_beam} is invalid.

\section{Anisotropic medium satisfying $r_s = r_p$ and $\dot{r}_s = \dot{r}_p$}
This section demonstrates that an interface between isotropic and anisotropic media with a specific condition supports $r_s = r_p$ and $\dot{r}_s = \dot{r}_p$ at any $\theta_i$, and these equalities yield an incident-polarization-independent SHEL at all $\theta_i$. We also show that a relaxed condition can yield the polarization-independent SHEL at near-normal incidence.

\begin{figure}[h!] \centering
	\includegraphics[width = \textwidth]{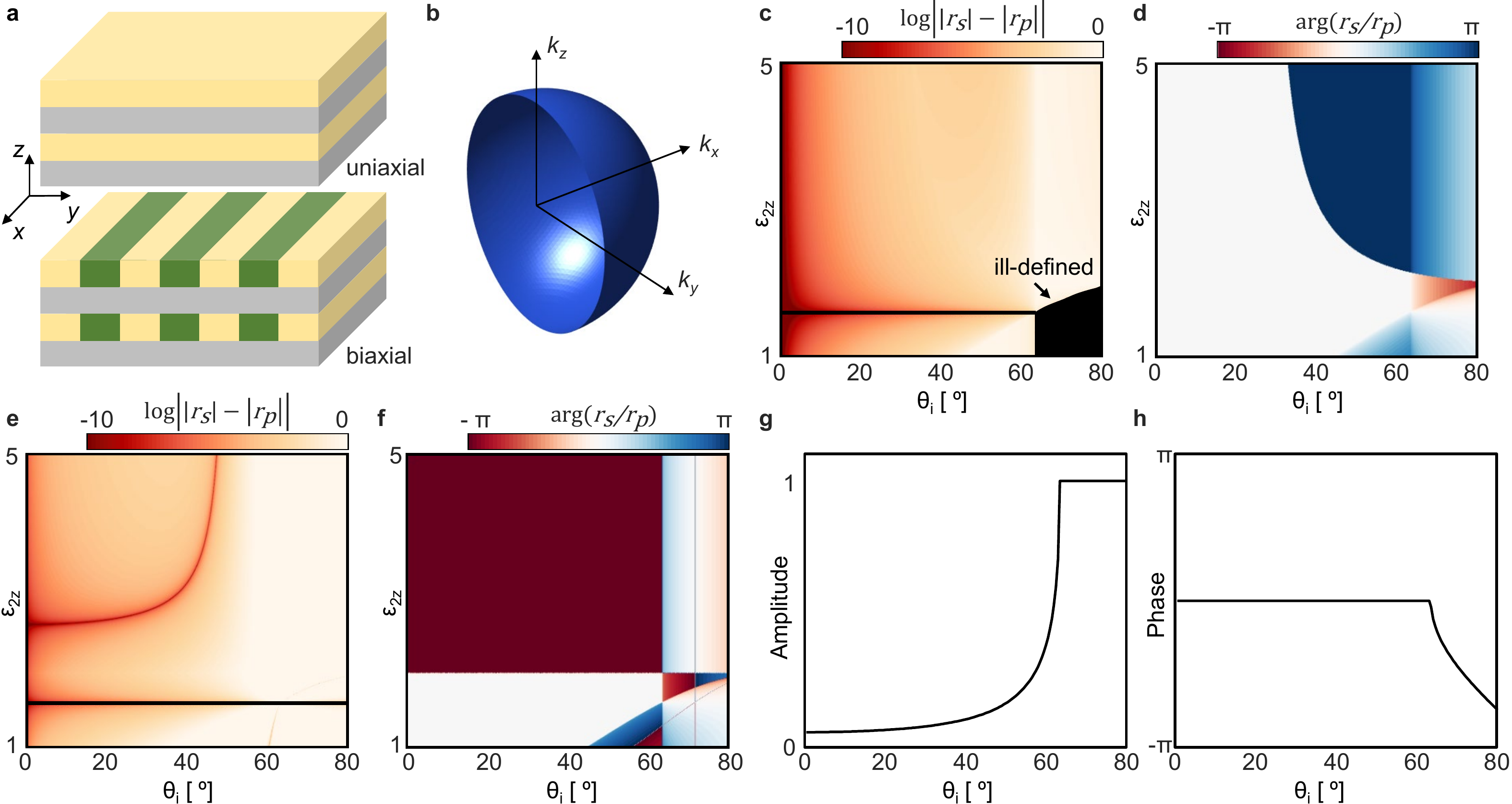}
	\caption{Optical properties of an anisotropic medium ($\varepsilon_2 = \text{diag}(\varepsilon_{2x}, \varepsilon_{2y}, \varepsilon_{2z})$). (a) Schematic of the anisotropic medium consisting of isotropic materials. (Top) Uniaxial anisotropic medium composed of two isotropic dielectrics and (bottom) biaxial anisotropic medium composed of three isotropic dielectrics. (b) A half of isofrequency contour ($k_x > 0$) of the anisotropic medium for $\varepsilon_{2x} = 2.5, \varepsilon_{2y} = \varepsilon_{2z} = 1.6$. (c) Amplitude and (d) phase difference of $r_s$ and $r_p$ at an interface between an isotropic dielectric ($\varepsilon_1 = 2$) and the anisotropic medium ($\varepsilon_{2x} = 2.5$ and $\varepsilon_{2y} = 1.6$) for various $\theta_i$ and $\varepsilon_{2z}$. The amplitude difference on a logarithmic scale is ill-defined in the black-shaded area (where $\lvert r_s \rvert = \lvert r_p \rvert$). (e) Amplitude and (f) phase difference of $\dot{r}_s$ and $\dot{r}_p$ at the interface. (g) Amplitude and (h) phase of $r_s$ and $r_p$ when $\varepsilon_{2z} = 1.6$. The two reflection coefficients are equal in the whole range and thus appear as a single curve.}
	\label{anisotropic}
\end{figure}

Two linear polarizations, $s$ and $p$, are degenerate at an interface between two isotropic media under normal incidence. Thus, the interface only supports polarization-independent reflection coefficients $r_s = -r_p$ at $\theta_i = 0^\circ$, where the minus sign is attributed to the sign convention in Fresnel equations \cite{born2013principles}. To increase the degrees of freedom, we examine a low-symmetry case of an anisotropic medium ($\varepsilon_2 = \text{diag}(\varepsilon_{2x}, \varepsilon_{2y}, \varepsilon_{2z})$). If two of the three permittivities are the same, the medium is uniaxial and can be realized by combining two isotropic dielectric materials (Fig. \ref{anisotropic}a, top). For the most general case of $\varepsilon_{2x} \neq \varepsilon_{2y} \neq \varepsilon_{2z}$, the medium is biaxial and can be realized by stacking different isotropic dielectrics (Fig. \ref{anisotropic}a, bottom). The isofrequency contour of the medium follows
\begin{equation}
	\frac{1}{k_0^2}\Big(\frac{k_x^2 + k_y^2}{\varepsilon_{2z}} + \frac{k_x^2 + k_z^2}{\varepsilon_{2y}} + \frac{k_y^2 + k_z^2}{\varepsilon_{2x}}\Big) - \Big(\frac{k_x^2}{\varepsilon_{2y} \varepsilon_{2z}} + \frac{k_y^2}{\varepsilon_{2x} \varepsilon_{2z}} + \frac{k_z^2}{\varepsilon_{2x} \varepsilon_{2y}} \Big) \frac{k_x^2 + k_y^2+k_z^2}{k_0^4} = 1,
\end{equation}
half of which ($k_x > 0$) is represented in Fig. \ref{anisotropic}b. The Fresnel coefficients at an interface between the anisotropic and isotropic ($\varepsilon_1$) medium can be represented as
\begin{align}
	r_s =& \frac{\sqrt{\varepsilon_1 - \beta^2} - \sqrt{\varepsilon_{2y}- \beta^2}}{\sqrt{\varepsilon_1 + \beta^2} + \sqrt{\varepsilon_{2y} - \beta^2}}, \notag \\
	r_p =& \frac{\sqrt{\varepsilon_1 - \beta^2} /\varepsilon_1- \sqrt{\varepsilon_{2x} - \beta^2\varepsilon_{2x}/\varepsilon_{2z}}/\varepsilon_{2x}}{\sqrt{\varepsilon_1 - \beta^2} /\varepsilon_1+ \sqrt{\varepsilon_{2x}-\beta^2 \varepsilon_{2x}/\varepsilon_{2z}}/\varepsilon_{2x}}
	\label{Fresnel}
\end{align}
where $\beta = \sqrt{\varepsilon_1}\sin\theta_i$ is the propagation constant. Eq. \ref{Fresnel} shows that $r_s = r_p$ for any $\theta_i$ if
\begin{equation}
	\varepsilon_{2x} \varepsilon_{2y} = \varepsilon_{2x} \varepsilon_{2z} = \varepsilon_1^2.
	\label{cond}
\end{equation}
This condition does not require the use of dispersive materials and can be readily fulfilled with positive permittivities by combining several dielectric materials (Fig. \ref{anisotropic}a, see Supporting information S1).

To investigate the Fresnel coefficients for different permittivities numerically, amplitude difference on a logarithmic scale (Fig. \ref{anisotropic}c) and phase difference (Fig. \ref{anisotropic}d) of $r_s$ and $r_p$ when $\varepsilon_1 = 2, \varepsilon_{2x} = 2.5$, and $\varepsilon_{2y} = 1.6$ are obtained. Indeed, the condition $r_s = r_p$ is satisfied at all $\theta_i$ when $\varepsilon_{2z} = 1.6$, where Eq. \ref{cond} is satisfied. The equality at all $\theta_i$ automatically imposes the equality of their derivatives, $\dot{r}_s = \dot{r}_p$ (Fig. \ref{anisotropic}e and f). Amplitude and phase of $r_s$ and $r_p$ when $\varepsilon_{2z} = 1.6$ demonstrate clearly that the Fresnel coefficients are equal and thus the two conditions $r_s = r_p$ and $\dot{r}_s = \dot{r}_p$ are satisfied in the entire range of $\theta_i$ (Fig. \ref{anisotropic}g and h). 

Furthermore, the two conditions of $r_s = r_p$ and $\dot{r}_s = \dot{r}_p$ are also satisfied under normal incidence for any $\varepsilon_{2z}$ if $\varepsilon_{2x} \varepsilon_{2y} = \varepsilon_1^2$ is satisfied (Fig. \ref{anisotropic}c-f) for the following reasons. First, one can notice straightforwardly from Eq. \ref{Fresnel} that $\varepsilon_{2z}$ does not affect on $r_s$ or $r_p$ when $\theta_i = 0^\circ$. Thus, a medium that supports $\varepsilon_{2x} \varepsilon_{2y} = \varepsilon_1^2$ satisfies $r_s = r_p$ under normal incidence, regardless of $\varepsilon_{2z}$ (Fig. \ref{anisotropic}c and d). Secondly, taking the first-order derivative of Eq. \ref{Fresnel} with respect to $\theta_i$ proves that $\dot{r}_s = \dot{r}_p = 0$ for any permittivity at $\theta_i = 0^\circ$. Thus, $\dot{r}_s = \dot{r}_p$ holds automatically under normal incidence, which is also confirmed by numerical results (Fig. \ref{anisotropic}e and f). While this discussion assumes the normal incidence, in which the SHEL does not occur, both $r_s/r_p$ and $\dot{r}_s/\dot{r}_p$ are close to unity with extremely high accuracy at small $\theta_i$ (Fig. \ref{anisotropic}c-f, see supporting information S2). Thus, the incident-polarization-independent SHEL is expected at a sufficiently small $\theta_i$ if $\varepsilon_{2x} \varepsilon_{2z} \neq \varepsilon_{2x} \varepsilon_{2y} = \varepsilon_1^2$.

\section{The polarization-independent SHEL under arbitrarily polarized incidence}
\begin{figure}[h!] \centering
	\includegraphics[width = \textwidth]{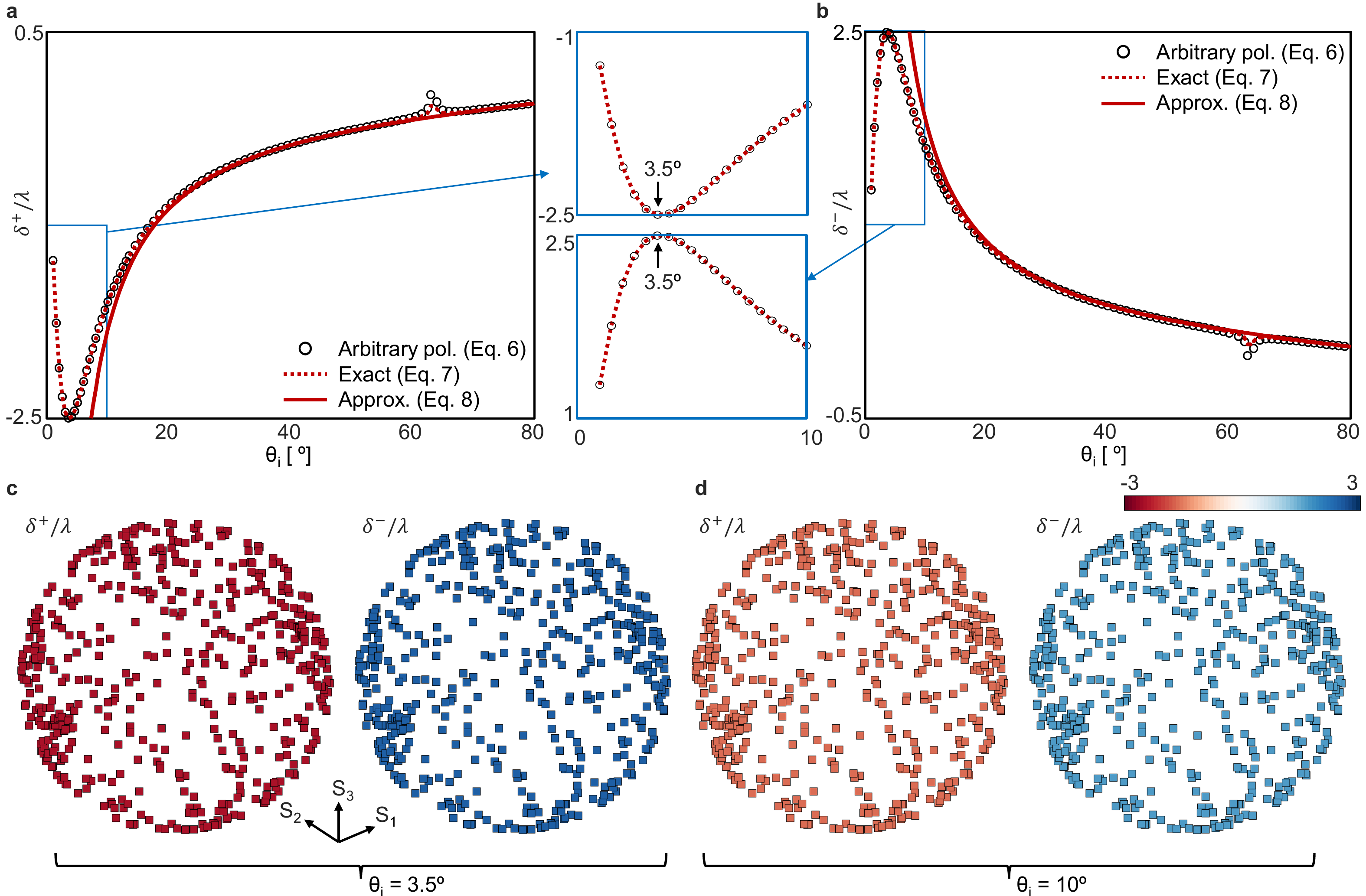}
	\caption{The spin Hall shift of $N$ number of arbitrarily polarized incidences. (a) $\delta^+/\lambda$ and (b) $\delta^-/\lambda$ under $N$ incidences calculated by Eq. \ref{yaverage}. All $\delta/\lambda$ have the same degenerate value and appear as a single marker at each $\theta_i$. Dashed and solid curves correspond to $\delta/\lambda$ obtained from the exact (Eq. \ref{shift_exact}) and approximated (Eq. \ref{shift}) formulas respectively. Plots in blue boxes show a magnified view at $\theta_i \leq 10^\circ$. (c, d) $\delta^\pm/\lambda$ represented in Poincar\'e sphere when (c) $\theta_i = 3.5^\circ$ and (d) $\theta_i = 10^\circ$.}
	\label{monte}
\end{figure}

To verify that the spin Hall shifts under arbitrarily polarized incidences are degenerate even when the large beam waist condition (Eq. \ref{large_beam}) does not hold, the SHEL at the interface that satisfies Eq. \ref{cond} ($\varepsilon_1 = 2, \varepsilon_{2x} = 2.5, \varepsilon_{2y} = \varepsilon_{2z} = 1.6$, Fig. \ref{anisotropic}g and h) is considered under $N = 500$ incidences, whose polarization states are randomly distributed over the Poincar\'e sphere (Fig. \ref{monte}). Each incidence has randomly assigned Stokes parameters $(S_1, S_2, S_3)$ that obey $\sum_{i=1}^3 S_i^2 = 1$. In calculations, we use $w_0 = 5\lambda$ to break the large beam waist condition (Eq. \ref{large_beam}) at small $\theta_i$ ($< 20^\circ$). In this tightly-confined regime, $\delta^\pm/\lambda$ calculated by the exact (Eq. \ref{shift_exact}) and approximated (Eq. \ref{shift}) formulas deviate from each other (Fig. \ref{monte}a and b, dashed and solid curves respectively). Whereas the polarization-dependent spin Hall shifts under these $N$ arbitrarily polarized incidences have distinct values at a given $\theta_i$ and are scattered around \cite{https://doi.org/10.1002/lpor.202100138} (see supporting information S2), the shifts here are independent of the incident polarization in the entire range and appear as a single marker at each $\theta_i$. In particular, at $\theta_i = 3.5^\circ$, the polarization-independent spin Hall shift reaches $\pm w_0/2$ (Fig. \ref{monte}a and b, insets), which is the maximum available value \cite{miao2021limitations}. In short, a gigantic SHEL that is comparable to $w_0$ and reaches the theoretical upper limit occurs independent of polarization. Small peaks near $\theta_i = 60^\circ$ (Fig. \ref{monte}a and b) originate from the diverging $\dot{r}_s$ and $\dot{r}_p$ at a critical angle (Fig. \ref{anisotropic}g and h).

For completeness, the spin Hall shifts under $N$ incidences are represented on the Poincar\'e sphere for two different angles, $\theta_i = 3.5^\circ$ (Fig. \ref{monte}c) and $\theta_i = 10^\circ$ (Fig. \ref{monte}d). The spin Hall shifts are degenerate to the result of the exact formula (Eq. \ref{shift_exact}) regardless of the Stokes parameters of the incidence. Naturally, the standard deviation is zero at all $\theta_i$ and is not shown here. In contrast, an interface with $\varepsilon_{2x} \varepsilon_{2z} \neq \varepsilon_{2x} \varepsilon_{2y} = \varepsilon_1^2$ satisfies $r_s = r_p$ and $\dot{r}_s = \dot{r}_p$ only under normal incidence and hence yields the polarization-insensitive SHEL at a small $\theta_i$ ($< 5^\circ$, see Supporting information S2). Considering that the large SHEL appears at the small $\theta_i$, this relaxed condition can be a promising alternative to realize polarization-independent spin Hall shift in a beam waist scale.

\section{The large SHEL reaching a half beam waist and its demonstration as a beam splitter}
\begin{figure}[h!] \centering
	\includegraphics[width = \textwidth]{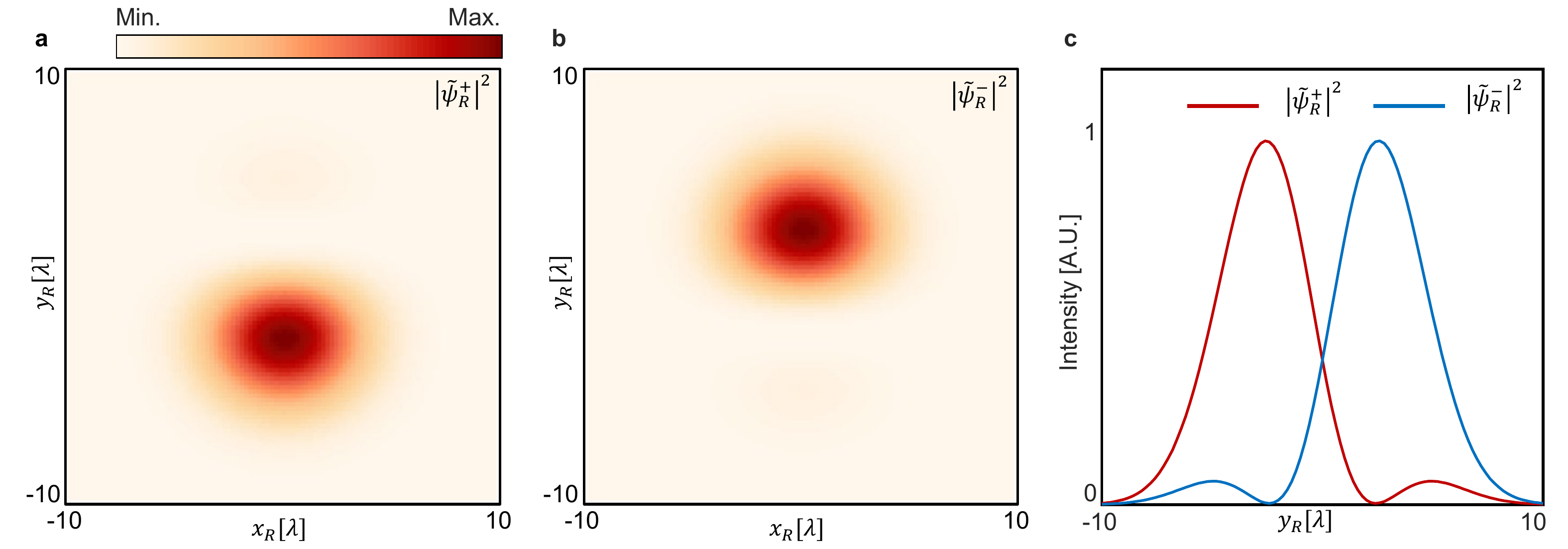}
	\caption{A discernible SHEL reaching a half beam waist under unpolarized incidence at $\theta_i = 3.5^\circ$. Intensity profiles of (a) LCP and (b) RCP components. (c) A cross-sectional view of intensities along the $y_R$-axis.}
	\label{field}
\end{figure}

To demonstrate the huge spin Hall shift that reaches $w_0/2$, the spatial field distributions of the circularly polarized components of the reflected beams are examined under unpolarized incidence (Fig. \ref{field}). Whereas the spin Hall shift is generally much smaller than $w_0$ and is not manifested visually \cite{kim2019observation, https://doi.org/10.1002/lpor.202100138}, here the spin Hall shift along the opposite directions of the $y_R$-axis is clearly distinguished (Fig. \ref{field}a and b). A cross-sectional view along the transverse axis demonstrates perceptible spin-dependent splitting (Fig. \ref{field}c). A weak side lobe centered at the opposite direction (approximately at $-5\lambda$) is observed for each LCP and RCP component. Between the side lobe and the maximum peak, the first minimum of the beam shape is located at $\pm 2.4\lambda$, which is closer to the center than the maximum of the oppositely circularly polarized beam is. Thus, according to the Rayleigh criterion, the splitting is resolvable when examined by the whole intensity. It is worth to emphasize that this large SHEL originates from the violation of the large beam waist condition (Eq. \ref{large_beam}), and hence the beam-waist-scale, polarization-insensitive SHEL cannot be realized using the previous condition of $r_s = r_p$ only \cite{https://doi.org/10.1002/lpor.202100138}.

\begin{figure}[h!] \centering
	\includegraphics[width = \textwidth]{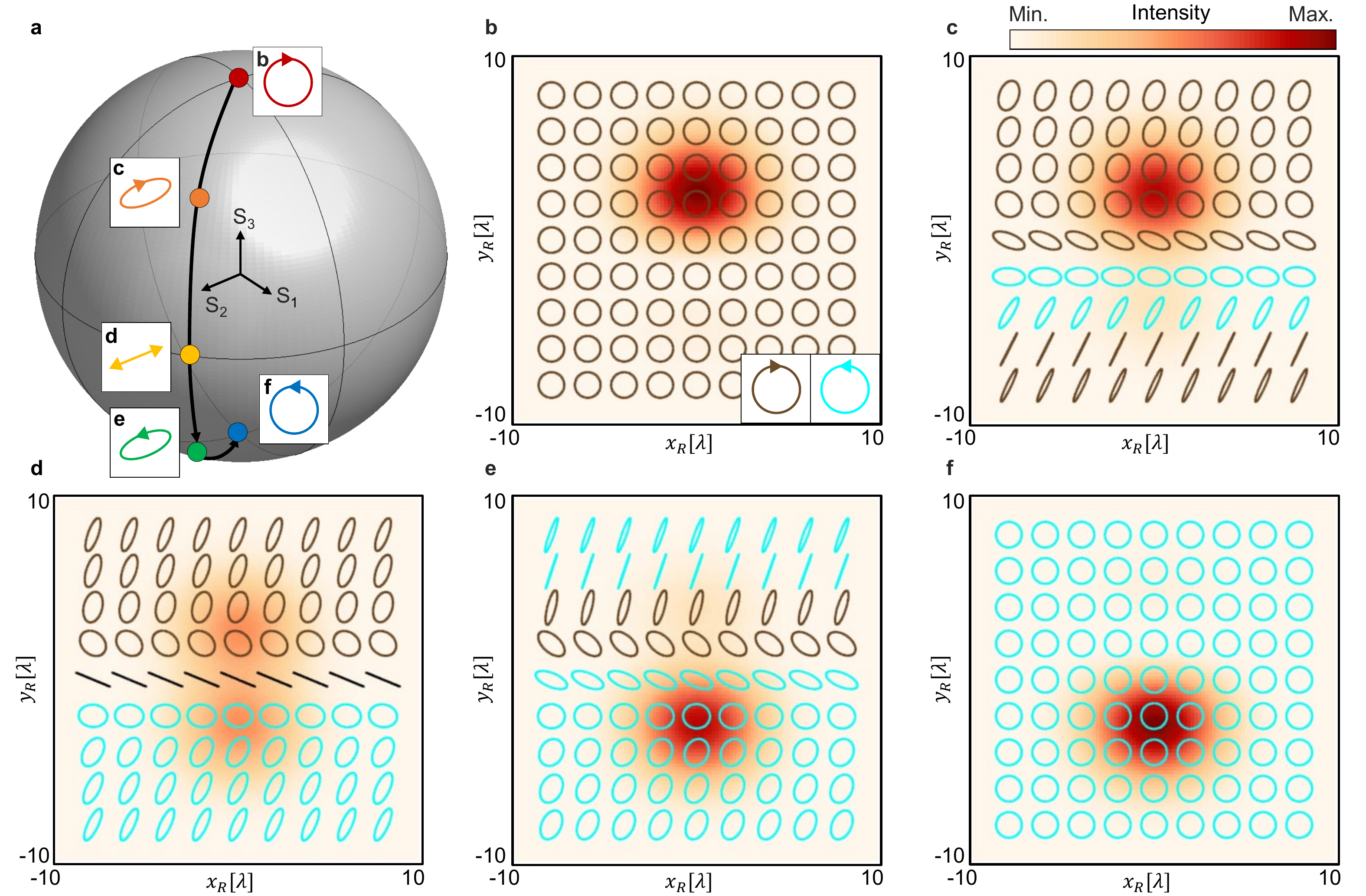}
	\caption{The SHEL as a beam splitter at $\theta_i = 3.5^\circ$. (a) Poincar\'e sphere with five polarization states marked. (b-f) The intensity profiles and local polarization states of the reflected beam under incidences whose polarization states are represented as (b) red, (c) orange, (d) yellow, (e) green, and (f) blue markers. Cyan and brown correspond to the left and right handedness respectively. Black denotes the linear polarization.}
	\label{intensity}
\end{figure}
Interestingly, the spin Hall shift is degenerate under all polarization states, but the intensity ratio between the LCP and RCP components is not. Thus, this SHEL can be exploited in intensity-tunable devices by modulating the incident polarization states. To prove this point, the intensity distribution of the reflected beam is examined for five different incident polarizations (Fig. \ref{intensity}). Firstly, in contrast to polarization-dependent SHEL, in which the direction and magnitude of the splitting change as the incident polarization varies, incidences with different polarization states produce the same spin Hall shift (Fig. \ref{large_beam}b-f). Meanwhile, the intensity splitting between LCP and RCP components directly follows the handedness of the incidence, i.e., its $S_3$. More specifically, the intensities of the circularly polarized components are associated with the third Stokes parameter of the incidence as $\int{\lvert \psi_R^\pm \vert^2 d\vec{r}} = (1 \mp S_3)/2$. Under circularly polarized incidence, the entire reflected beam undergoes a displacement without being split (Fig. \ref{intensity}b and f). In contrast, the reflected beam is split into two peaks with inequal intensities when the incidence is elliptically polarized (Fig. \ref{intensity}c and e). Under linearly polarized incidence, two equally-split peaks, each for LCP and RCP, appear. Whereas the SHEL that is much smaller than the beam waist can be resolved only through amplification, the two peaks are identifiable (Fig. \ref{intensity}d). The linear polarization state at the center ($y_R = 0$) is mirror-symmetric to the incident polarization with respect to the $y_R$-axis as a result of the $\pi$-shifted phase of $r_p$.

We demonstrate the polarization-independent SHEL of the tightly-confined beam where $w_0$ is comparable to $\lambda$ as a proof-of-concept. However, it should be noted that a beam that has $w_0 \gg \lambda$ can also violate the large beam waist condition (Eq. \ref{large_beam}) at a sufficiently small $\theta_i$ ($\ll 1^\circ$) and our theory is also applicable to this regime. In such a case, the upper limit of the spin Hall shift ($w_0/2$) becomes even greater as a result of the large $w_0$, but the SHEL requires a precise setting of the ultrasmall $\theta_i$ (see supporting information S3).

\newpage
\section{Conclusion}
In conclusion, an approach towards the incident-polarization-insensitive spin Hall effect of light that exhibits a discernible splitting is presented. A theoretical proof of the required conditions, $r_s = r_p$ and $\dot{r}_s = \dot{r}_p$, is given, followed by a suggestion of an interface that supports such conditions in the entire range of incident angles. Also, a relaxed condition that yields a large and polarization-independent spin Hall effect of light at a small incident angle is presented. In addition, we demonstrate the polarization independence of the spin Hall effect of light using Monte-Carlo simulations under a huge number of arbitrarily polarized incidences. The gigantic and polarization-independent splitting that is comparable to the beam waist proves potential of our work to enable spin-dependent devices that are invariable under the change of incident polarization at any angle and active splitters that act upon incident polarization.




\newpage
\bibliographystyle{custom}
\bibliography{reference}

\end{document}